\documentclass[12pt,preprint]{aastex}

\newcommand{\Draft}{false}

\usepackage{amsmath}
\usepackage{amssymb}
\usepackage{xspace}
\usepackage{color}
\usepackage{ifthen}
\usepackage{xspace}
\usepackage{natbib}

\definecolor{gray}{rgb}{0.5,0.5,0.5}

\newcommand{\lSect}[1]{{\label{sec:#1}}}
\newcommand{\lFig}[1]{{\label{fig:#1}}}

\newcommand{\Sectff}[1]{{\ref{sec:#1}}}
\newcommand{\Sect}[1]{{\S~\Sectff{#1}}}

\def\gtaprx {\lower .1ex\hbox{\rlap{\raise .6ex\hbox{\hskip .3ex
	{\ifmmode{\scriptscriptstyle >}\else
		{$\scriptscriptstyle >$}\fi}}}
	\kern -.4ex{\ifmmode{\scriptscriptstyle \sim}\else
		{$\scriptscriptstyle\sim$}\fi}}}
\def\ltaprx {\lower .1ex\hbox{\rlap{\raise .6ex\hbox{\hskip .3ex
	{\ifmmode{\scriptscriptstyle <}\else
		{$\scriptscriptstyle <$}\fi}}}
	\kern -.4ex{\ifmmode{\scriptscriptstyle \sim}\else
		{$\scriptscriptstyle\sim$}\fi}}}

\slugcomment{Draft for ApJ, \today} 

\shorttitle{The Progenitor Stars of Gamma-Ray Bursts}
\shortauthors{Woosley and Heger}

\begin{document}


\title{The Progenitor Stars of Gamma-Ray Bursts}


\author{S.~E.\ Woosley\altaffilmark{1}, A.~Heger\altaffilmark{2}}

\altaffiltext{1}{Department of Astronomy and Astrophysics,
University of California, Santa Cruz, CA 95064; woosley@ucolick.org, cumming@ucolick.org}
\altaffiltext{2}{Theoretical Astrophysics Group, MS B227, Los Alamos National Laboratory, Los
Alamos, NM 87545; 1@2sn.org}

\begin{abstract} 
Those massive stars that, during their deaths, give rise to gamma-ray
bursts (GRBs) must be endowed with an unusually large amount of
angular momentum in their inner regions, one to two orders of
magnitude greater than the ones that make common pulsars. Yet the
inclusion of mass loss and angular momentum transport by magnetic
torques during the precollapse evolution is known to sap the core of
the necessary rotation. Here we explore the evolution of very rapidly
rotating, massive stars, including stripped down helium cores that
might result from mergers or mass transfer in a binary, and single
stars that rotate unusually rapidly on the main sequence.  For the
highest possible rotation rates (about 400 km s$^{-1}$), a novel sort
of evolution is encountered in which single stars mix completely on
the main sequence, never becoming red giants. Such stars, essentially
massive ``blue stragglers'', produce helium-oxygen cores that rotate
unusually rapidly. Such stars might comprise roughly 1\% of all stars
above 10 solar masses and can, under certain circumstances retain
enough angular momentum to make GRBs. Because this possibility is very
sensitive to mass loss, GRBs will be much more probable in regions of
low metallicity.
\end{abstract}


\section{Introduction}
\lSect{intro}

During the last seven years, compelling evidence has accumulated that
gamma-ray bursts (GRBs) of the ``long-soft'' variety are a consequence
of the deaths of massive stars. GRBs occur in galaxies and regions of
galaxies where vigorous star formation is going on (Vreeswijk et al
2001; Fruchter et al 1999; Fruchter et al, 2005, in preparation, as
cited in Levan et al 2005). Some, perhaps even a large fraction, are
accompanied by supernovae of an unusual kind \citep{Gal98,Hjo02,Sta03,
Zeh04,Lev05}. These supernovae have lost their hydrogen envelopes and
are observed to be Type Ibc. This is consistent with the theoretical
expectation that even a relativistic jet cannot escape a blue or red
supergiant with enough energy to make a GRB that lasts only tens of
seconds \citep{Woo93,Mac99,Zha04}. It is clear therefore that at least
some GRBs are made when massive Wolf-Rayet (WR) stars die. It is
equally clear that not all WR star deaths make GRBs.

The discriminating characteristics of those stars that do make GRBs
are very likely their mass and rotation rate. All currently favored
models for GRBs require so much rotation that it plays a dominant role
in the explosion mechanism. This is true of the collapsar model
\citep{Mac99} where enough angular momentum must be present to form
disk around a black hole of several solar masses. It is also true of
the ``millisecond magnetar'' model \citep{Uso94,Whe00}, where the
rotational energy must be sufficient to produce magnetic fields of
order 10$^{16}$ gauss, and of the ``supranova'' model \citep{Vie98}
where a neutron star supported, in part, by rotation must exist for an
extended period. The energies inferred for the ``hypernovae''
associated with GRBs \citep{Iwa98,Woo99,Den05} are $\sim10^{52}$
erg. If the energy is to be derived from rotation, the corresponding
pulsar, in those models that use pulsars, must have an initial
rotation rate of $\ltaprx1.5$ ms. This is more an order of magnitude
faster than the fastest observed pulsars and close to what a neutron
star can tolerate without deformation or excluding a disk. Collapsar
models need even more rotation and are favored by high stellar mass,
which more readily gives the requisite black hole. So, some GRBs are
the violent deaths of massive WR-stars whose cores are very rapidly
rotating.

This rapid rotation of the inner core is at variance with what is
needed to make common pulsars with rotation periods $\gtaprx15$ ms
\citep{Heg05}. There are two possibilities. Iron cores generally
collapse with high rotation rates, but the rotation is damped during
or shortly after the explosion either by gravitational radiation or by
the transport of angular momentum to the ejecta. Or the core is
already rotating slowly enough that its angular momentum does not
exceed that of a typical pulsar ($\sim 5 \times 10^{47}$ erg s). The
first possibility does not seem likely at the present
time. Gravitational radiation by the $r$-mode instability 
does not slow a 1 ms pulsar to a period of 10 ms in a few centuries
\citep{Arr03} and the dissipation of the rotational energy of a 1 ms
pulsar by means other than gravity waves or neutrinos would give much
more energy than is observed in a typical supernova.  These are not
iron clad arguments, but do motivate the study of the second
possibility.

That too leads to a conundrum. If typical massive star death gives
slow pulsars, what special circumstances give a GRB? In \citet{Heg05},
we showed that current estimates of magnetic torques in the interiors
of evolving massive stars \citep{Spr02} led naturally to rotation
rates of pulsars in the range 10 - 15 ms, just what was needed. But
this was for the most common variety of supernovae, Type IIp, that
result from the deaths of red supergiants (RSGs). More massive stars
and especially stars that lost their hydrogen envelopes early on, had
cores that rotated more rapidly. Could it be that some fraction of
those massive stars that evolve through a giant phase and lose their
envelopes either to winds or companion stars end up making GRBs. The
answer is ``maybe'', but it's not easy. There are the twin problems of
magnetic torques and WR-mass loss. Without compelling reasons to the
contrary, one must employ the same prescription for magnetic torques
in the evolution of GRB progenitors as for pulsar progenitors, and the
core spins down a lot during the RSG phase. Second, even if the
envelope is removed, the vigorous mass loss of typical WR stars still
carries away a lot of angular momentum.

It has been known for some time that if the magnetic torques are
negligible, which is to say much weaker than estimated by
\citet{Spr02}, it is easy to give GRB progenitors the necessary
angular momentum \citep{Heg00, Hir04,Hir05}, but then one must
invoke another mechanism to slow down typical pulsars. It has also
been known that if the magnetic torques are included and standard mass
loss rates are employed, that it is very difficult to make GRBs
\citep{Heg03,Woo04} from any star that either passes through a giant
phase or loses appreciable mass as a WR star.
  
Here we consider possible resolutions to this dilemma.  All of them
require rapidly rotating stars to begin with and a decrease of up to a
factor of 10 in the standard WR-mass loss rates currently in use by
the community. This may not be so great an adjustment as it
sounds. Not only does one expect some scatter in the mass loss rates
of stars having the same mass, composition and angular momentum on the
main sequence, but WR stars of lower metallicity are known to have
lower mass loss rates \citep{Vin05}. Moreover, the rate of mass loss
may not necessarily equal the (angle-averaged) rate of angular
momentum loss \citep{Mae02}, since the mass loss could occur
predominantly at the poles.

We consider two possibilities, simple helium cores parameterized by
their rotational speeds (some fraction of break up) at helium ignition
and rapidly rotating single stars that experience complete mixing on
the main sequence. The former set of stars might be illustrative of
the outcome of stellar mergers or other binary activity
\citep{Sma02,Pod04,Fry05,Pet05}.  The latter represents, so far as we
know, a novel suggestion for the evolution of GRB progenitors.

\section{Models and Physics}
\lSect{models}

The stars are evolved using the KEPLER implicit hydrodynamics package
\citep{Wea78} using the same parameters for, and treatment of angular
momentum transport and mixing as in \citet{Heg00} and
\citet{Heg05}. Additional discussion of the physics is given in those
references and in \citet{Pet05}. For those calculations that include
magnetic torques, the formalism of \cite{Spr02} is employed.

A major deficiency of the code is that it is one dimensional, so any
deformation due to rotation is not followed. All rotational quantities
such as angular momentum, torques, etc. are angle averaged, and the
Lagrangian mass shells transport these average quantities. The
rotation is thus ``shellular''. An even greater deficiency is that
rotation is treated as a passive quantity with no back reaction on the
stellar structure, that is the centrifugal term is not included in the
force equation. This is not a bad approximation so long as the ratio
of centrifugal force to gravity remains small. In practice, this is
true except for the outermost layers at the end of helium
burning. However, in the most rapidly rotating stars studied here,
centrifugal force can approach or even mildly exceed unity in a small
fraction of the mass (\Sect{runs}).  It is likely that these layers
are ejected to form a disk around the star, but our present treatment
is unable to follow this realistically.

\subsection{Rotation Rate}
\lSect{rotate}

Main sequence stars of type O and B are known to be rapid rotators.
About 0.3\% of B stars have rotational speeds on the main sequence in
excess of two-thirds times the break up speed, and the average is 25\%
of break up \citep{Abt02}.  The fastest rotating stars considered here
have about 45\% of break up, or rotational speeds on the main sequence
in excess of 350 km s$^{-1}$. \citet{Gie05} estimate that this would
be several per cent of field B stars. This is consistent with our
expectation that only a percent or so of stars over 10 M$_{\sun}$ have
the special properties required to make a GRB.

The rotation rate for WR stars is not well determined observationally,
but is expected, on theoretical grounds, to be rapid, at least for low
metallicity \citep{Mey05}. We are interested here in a small fraction
of WR stars that may have experienced unusual evolution and thus feel
justified in assuming large values up to and including those that
would cause large deformation.

\subsection{Mass Loss and Angular Momentum Loss}
\lSect{mdot}

Our results will be quite sensitive to the rates adopted for mass
loss, especially for hydrogen-deficient stars.  For main sequence
stars and red supergiants the mass loss rates employed in these
studies were taken from \citet{NJ90}.  For WR stars, a mass-dependent
mass loss rate \citep{Lan89} was assumed using the scaling law
established by \citet{WL99}, but lowered by a factor 3 \citep{HK98} to
account for clumping \citep{Nu98}, though see \citet{Bro04}.
Wind-driven mass loss in main sequence stars is believed to be
metallicity dependent and a scaling law $\propto\sqrt{Z}$ has also
been suggested for hot stars \citep{Kud00,Kud02,NL00,Cro02}.
\citet{WHW02} assumed that the same scaling law holds for WR stars
\citep{Van01} and blue and red supergiants as well.  ``Metallicity''
is assumed here to be the initial abundance of heavy elements,
especially of iron, not the abundances of new heavy elements like
carbon and oxygen in the atmospheres of WC and WO stars. The situation
has been recently examined for WR stars of Type WN and WC by
\citet{Vin05} who find a scaling law $\propto Z^{0.86}$ for
metallicities in the range log $Z/Z_{\odot}$ = -1 to 0, with a more
gradual decline below -1 for WC stars, but continuing to at least -2
for WN stars. Since our stars spend significant fraction of their
lifetimes as WN stars and since we consider stars only down to
metallicity -2, this revised scaling suggests that the winds we
actually employed in the present study could have been smaller by a
factor of $\sim Z^{1/3}$. This would make the production of GRB
progenitors more likely, as we shall see.

It will turn out that we are also sensitive to the mass loss rates for
stars that, though still on the main sequence because their central
hydrogen abundances have not gone to zero, have low surface hydrogen
abundances due to deep rotational mixing. We have used WR-mass loss
rates for all stars with surface temperatures over 10,000 K and
hydrogen mass fraction less than 40\%. Such loss rates may
well be too large, but the difference in final mass between stars where
WR mass loss is not implemented until helium ignition is not large

The most important effect of mass loss in the present context is to
carry away angular momentum. It is generally assumed that the momentum
lost is just the angle-averaged value at the surface times the mass
loss rate. However, it may well be that precisely those very rapidly
rotating WR stars we want to consider have quite anisotropic mass
loss. The higher temperature and luminosity at the poles makes the
loss greater there. If this is a significant effect, \citet{Mae02} has
shown that the angular momentum loss might be considerably reduced,
perhaps even leading to ``breakup'' in OB stars.

Here we consider this possibility as just another uncertainty in the
mass loss rate which might again allow us to use values somewhat smaller
than the standard ones.  It should also be noted that the mass loss
rates for GRB progenitors inferred from their afterglows are generally
much smaller than the standard values. \citet{Sod05}, for example,
determine a pre-explosive mass loss rate for Type Ic SN 1998bw of $6
\times 10^{-6}$ M$_{\sun}$ yr$^{-1}$, even though the accepted mass
for the supernova is thought to have been around 10 M$_{\sun}$
\citep{Iwa98,Woo99}. The mass loss inferred for Type Ic SN 2002ap was $5 \times
10^{-7}$ M$_{\sun}$ yr$^{-1}$ \citep{Ber02}. We thus consider models
here that, in addition to the assumed scaling with metallicity, have
mass loss rates reduced by up to a factor of 10 from their standard
values scaled by $\sqrt{Z}$.

\subsection{Composition, Parameters, and the Naming of Models}
\lSect{params}

For the bare helium cores, a composition of 98.5\% He, 1.5\% $^{14}$N,
and a solar complement \citep{Lod03} of elements heavier than neon was
adapted. Since we are not interested in nucleosynthesis in this study
and since the energy generation rate and opacity are essentially
independent of the composition of trace elements throughout most of
the star, the composition enters in chiefly as a modification of the
assumed mass loss rate. The mass loss rate in these solar metallicity
models was scaled by factors of 1., 0.3, and 0.1 to reflect how things
might vary with metallicity and the uncertainties discussed in
\Sect{mdot}. A reduction factor of 10 for example, might correspond
either to a very metal deficient star of 1\% Z$_{\sun}$ (assuming mass
loss scales as $Z^{0.5}$) or to a star of 10\% $Z_{\sun}$ that had,
for whatever reason, an angular momentum loss three times smaller than
given by the fitting formula. For helium stars, we considered only a
single mass, 16 M$_{\sun}$, which is rather typical, after some mass
loss, to what has been discussed as a progenitor for the supernovae
seen with GRBs. Very similar results would characterize helium cores
that were within a factor of two of 16. The models are defined in
Table 1 by the ratio of centrifugal force to gravity at the surface at
the equator when the star has produced a central carbon mass fraction
of 1\%. Here that ratio is taken to be a large value, 47\%, but the
ratio decreases to 30\% just 0.25 M$_{\sun}$ into the star. The
surface rotational speed was 800 km s$^{-1}$. An additional set of
models with the initial angular momentum reduced by half, i.e., a
surface break up fraction of 26\%, gave essentially the same
presupernova core properties (Table 1).

For single stars, several choices of mass (12, 16, 35 M$_{\sun}$) and
metallicity (100\%, 10\%, 1\% Z$_{\sun}$) were explored (Table 2). The
name of each model is given by its zero age main sequence mass, a
letter indicating metallicity (``T'' = 1\% solar; ``O'' = 10\% solar;
and ``S''= solar), and another letter to distinguish choices for mass
loss rate during the WR stage and whether magnetic torques were
applied.

\section{Results and Discussion}
\lSect{runs}

Our principal results are given in Tables 1 and 2.  For the bare
helium cores that did not lose a lot of mass (Table 1), the equatorial
rotation rate remained a substantial fraction of critical throughout
the evolution. The maximum specific angular momentum and slowest
rotational period was always at the surface. This makes it difficult
to envision how any binary interaction could have imparted a faster
rotation to the inner core than what is calculated here.

\subsection{Helium Cores}
\lSect{hecore}

Helium cores that lost little mass (e.g., Models A, F, G, H, I, N, O,
P) remained nearly rigidly corotating throughout helium burning and
retained a nearly constant angular momentum. It is interesting to
compare Models HE16I and HE16P which, aside from magnetic torques,
were otherwise identical models.  We focus on the inner 3 M$_{\sun}$
because that is where the possibility of a millisecond pulsar or an
extreme Kerr black hole will be determined. Because there was
essentially no mass loss and the coupling between the inner and outer
core was not strong in Model HE16I, angular momentum was conserved in
the inner core throughout the evolution. At helium ignition, the
angular momentum inside 3 M$_{\sun}$ was $2.86 \times 10^{50}$ erg
sec; at the presupernova stage it was $2.78 \times 10^{50}$ erg sec.
The magnetic torques in HE16P on the other hand transferred
significant angular momentum out of the inner core. Both models had
the same angular momentum, to within a few percent, at helium
depletion (Y$_c$ = 0.01) and both were rigidly rotating.  Contraction
to a central temperature of $5 \times 10^8$ K also gave models whose
specific angular momenta differed by less than 10\% in the inner
core. However, from carbon ignition ($8 \times 10^8$ K) onwards they
diverged. By carbon depletion, the model without magnetic torques was
rotating three times faster in its inner core and by the time the iron
core collapsed this factor had become 8. Interestingly, the angular
momentum of the inner 3 M$_{\sun}$ of the model with magnetic torques
decreased by a factor of 2 after carbon depletion, that is during the
last 0.9 years of the stars life,

The evolution of the mass shedding helium cores differed in an
expectable way. The more mass lost, the slower was the rotation of the
outer core and, if magnetic torques were appreciable, the slower the
inner core as well. Consider Model HE16L which included magnetic
torques, had a mass loss rate 30\% that regarded as standard for solar
metallicity stars, and ended its life with a mass of 9.58 M$_{\sun}$.
Almost all the mass loss occurred during helium burning.  The angular
momentum in the inner 3 M$_{\sun}$ of its core was the same as the
other models at helium ignition ($3.32 \times 10^{50}$ erg sec) but by
helium depletion, it had declined by over an order of magnitude to
$2.79 \times 10^{49}$ erg sec. The entire star was still rotating
nearly rigidly. After carbon burning and the accompanying contraction
and spin up of the inner core, the angular momentum in the same 3
M$_{\sun}$ was reduced by an additional factor of 3 at carbon
depletion and an additional 30\% by the time of iron core collapse
($6.86 \times 10^{48}$ erg sec). At carbon depletion the spread in
angular velocity from center to surface varied by a factor of 5.

In general, one sees the tendency of magnetic torques to enforce rigid
rotation. This extracts angular momentum from the inner core when it
contracts and spins up in the post-helium burning stages. It also
brakes the core when extensive mass loss slows the rotation of the
outer layers.

In those models with appreciable mass loss the ratio of centrifugal
force to gravity decreases with time (Table 1) and is never greater
than the initial value. However, for models with little or no mass
loss and large magnetic torques, the centrifugal forces at the surface
increase during the latter stages of helium burning and carbon burning
to the point where they are comparable to gravity. This signals a
breakdown in our treatment of rotation, but fortunately occurs only in
the outer few hundredths (Models O and P) to few tenths (Models G and
H) of a solar mass. What probably happens here is that the star forms
a disk. Depending on the viscosity, some of that disk will reaccrete
so that one ends up with a star, roughly spherical throughout most of
its interior, but with highly deformed surface layers rotating at
break up in the equator. Whether this might have an adverse effect on
the rotation of the inner core is uncertain, but may have no more
effect than simply losing an equivalent amount of mass.

\subsection{Rapidly Rotating Single Stars}
\lSect{hstars}

In rapidly rotating single stars one has to deal with the additional
complexity of a hydrogen envelope. In models with typical rotation
rates, say 200 km s$^{-1}$, for the assumed mixing parameters, the
star becomes a supergiant sometime during helium burning. For stars
with very low metallicity, the supergiant may be blue (BSG), but
typically it is red (RSG) with a radius of several AU. The formation
of this extended, high mass, very slowly rotating envelope has a great
influence on the core, typically braking its rotation rate below that
required to make a GRB. In fact, the RSG branch of evolution was
extensively explored by \citet{Heg05} and, for the given
parameterization, was found to give typical periods inferred at birth
for pulsars.

The more rapidly rotating models, which are of greatest interest here,
bypass giant formation by remaining almost completely mixed on the
main sequence. Thus, at the end, they resemble the helium cores of the
previous section. An interesting exemplary case is Model 16TI. This
star rotated at about 400 km s$^{-1}$ on the main sequence and became
a WR star shortly after central H depletion. Because of its assumed
low metallicity, 1\% solar and mass loss scaling ($\dot M \propto
Z^{0.5}$), mass loss on the main sequence was negligible and mass loss
as a WR star was reduced by 10 compared to solar metallicity stars. In
this particular model the mass loss was decreased by an additional
factor of 3 to explore the consequences. The final mass of the 16
solar mass star was 13.95 M$_{\sun}$. It was a WO star with surface
abundances 40\% C, 40\%O, and 20\% He. Most of the presupernova star
was predominantly oxygen and heavier elements

Again focusing on the angular momentum in the inner 3 M$_{\sun}$ (see
also Fig. 1), the angular momentum when the star had burned about half
of its hydrogen (X$_c$ = 0.40 was $1.10 \times 10^{51}$ erg sec. This
declined by about 20\% at central hydrogen depletion and by about a
factor of 4 at helium ignition. This large decrease was caused by core
contraction and spin up coupled to the outer layers by magnetic
torques. However, half way though helium burning the angular momentum
in the inner 3 M$_{\sun}$ had increased back to $7.2 \times 10^{50}$
erg s owing to extensive convection redistributing angular momentum in
the core.  This decreased to $4.80 \times 10^{50}$ erg sec when the
central temperature was $5 \times 10^8$ K and to $1.38 \times 10^{50}$
erg sec at carbon depletion.  In marked contrast to the RSGs that make
pulsars \citep{Heg05}, the angular momentum in the inner core
continued to decline appreciably in the late stages of
evolution. After central carbon depletion, with only 0.16 years left
to live, the angular momentum in the inner 3 M$_{\sun}$ declined an
additional factor of almost 4 to $3.67 \times 10^{49}$ erg sec.  at
the presupernova stage. In the presupernova (central density $4.2
\times 10^9$ g cm$^{-3}$), the angular velocity in the inner solar mass
was 100 times that near the surface.

To summarize, for very rapidly rotating stars with magnetic torques
and little mass loss, angular momentum in the inner core is
essentially preserved throughout the main sequence and has only
declines by about a factor of 2 at helium depletion. The major angular
momentum loss occurs during carbon and oxygen burning with a large
fraction occurring during the final months of the star's life. This is
because of the very large differential rotation developed by the core
as it contracts through these advanced burning stages and the effect
of magnetic torques which try to maintain rigid rotation.

Like the helium cores, some of these rapidly rotating stars also
develop centrifugally supported surfaces. In the case of Model 16TI,
centrifugal forces exceed gravity at the equator during helium shell
burning, though only in the outer 0.1 M$_{\sun}$. Other models to
develop critical rotation after helium burning were Models 12TC, I and
J; 16TB, C, and J; 12OC, J, and N; and 16OC, J, M, and N. None of
the solar metallicity calculations or 35 M$_{\sun}$ stars developed
critical rotation.

\subsection{Massive Oe and  Be Stars?}
\lSect{Bestar}

Oe and Be stars are a subclass of massive stars that show emission
lines, usually taken to be indicative of a disk \citep{Han96}. There
is evidence for rapid rotation. Indeed Be stars are the most rapidly
rotating of all non-compact stars \citep{Tow04}.  These stars appear
may correspond to a phase of spin up caused by mass transfer in a
close binary system or by an internal redistribution of angular
momentum. Not all Be stars are observed in binary systems, so it is
possible that some form from single B stars.

The low metallicity, rapidly rotating stars considered here might
evolve through a stage having properties similar to Oe and Be
stars. However, this would only be for stars that had, for some
reason, very low mass loss rates. It is interesting though that the
stars which might develop disks are the same stars most likely to
produce GRBs. It may be that the precursor to a GRB is an Oe or Be
star.  However, the converse, that all Oe and Be stars make GRBs is
unlikely, especially in regions with solar metallicity.

\section{Presupernova Characteristics}
\lSect{presn}

The presupernova characteristics of the cores of our models are given
in Tables 1 and 2 and Fig. 2. Various entries give the baryonic mass
of the iron core that collapses, its total angular momentum, the
rotation rate a pulsar would have if the inner 1.7 M$_{\sun}$
collapsed and conserved angular momentum, and the Kerr parameter that
a black hole would have if it formed from the inner 3 M$_{\sun}$ of
the model. The baryonic mass of the iron core differs from the pulsar
mass for various reasons \citep{Tim96}, especially because of
accretion during the explosion and neutrino mass loss. The exact
relation is unknown because of uncertainties in the explosion
mechanism. However, a 1.7 M$_{\sun}$ (baryonic mass) core would give a
1.44 M$_{\sun}$ (gravitational mass) neutron star after neutrino
losses. Approximately 20\% could be added to the rotational period of
those models that give neutron stars and not black holes because of
the angular momentum carried away by the neutrinos \citep{Heg05}.

Stars that have an entry greater than 1 for the Kerr parameter at 3
solar masses would have to form a disk to carry the extra angular
momentum and are thus good candidates for collapsars. If a black hole
forms in these systems, so will a disk. Other models having $a \gtaprx
0.3$ at 3 M$_{\sun}$ are also good candidates because the angular
momentum increases outwards. Fig. 2 shows the angular momentum
distribution in Models 16TI and 16OM. Both would form
accretion disks at 3.5 and 5.5 M$_{\sun}$ respectively, even though
the Kerr parameter at 3 M$_{\sun}$ is only 0.44 and 0.25.

\section{Implications for Gamma-Ray Bursts and Supernovae}
\lSect{implications}

One can broadly characterize the effect of rotation on the explosion
mechanism by the rotational energy the resulting pulsar would have if
one formed and conserved angular momentum. For a typical neutron star
radius (12 km) and {\sl gravitational} mass (1.4 M$_{\sun}$), the
moment of inertia is $\approx 0.35 MR^2 = 1.4 \times 10^{45}$ g cm$^2$
\citep{Lat01}. The rotational energy, $1/2 I \omega^2$, is then
$E_{rot} \approx 1.1 \times 10^{51} \, (5 \ {\rm ms}/P)^2$ erg. Since
the typical kinetic energy of a supernova is 10$^{51}$ erg, this
implies a necessary condition that the pulsar contribute the bulk of
the energy is that its rotation rate be $\ltaprx$5 ms. This lays aside
all considerations of how this energy might be tapped and with what
efficiency. To give a ``hypernova'' with ten times this energy
requires rotational periods $\ltaprx$2 ms. It turns out that stars
that would give disks around black holes of several solar masses also
require a comparable rotation rate, $\ltaprx$1 ms, though the relevant
angular momentum is located somewhat farther out in the star.  So we
can make the distinction. Neutron stars with initial periods of 10 ms
or longer probably won't have a large effect on the explosion; those
with periods less than 5 ms might, and a 1 ms period is needed to make
a GRB.

By this criterion, all neutron stars resulting from stars that pass
through a supergiant phase (either red or blue) and do not lose their
envelopes will rotate too slowly to be GRBs, or even to power normal
supernovae. Even those stars resulting from helium cores rotating near
break up - whether formed from binary evolution, or very rapidly
rotating solitary stars - will be too slow to make GRBs {\sl unless
their mass loss rates are smaller than generally assumed}. This could
come about either because the metallicity of GRBs is quite low or
because the rates currently in use overestimate the actual angular
momentum loss for unknown reasons (\Sect{mdot}. The recent revision
downwards of mass loss rates for metal deficient WR stars by
\citet{Vin05} is helpful in this regard.

With reasonable variations then it is possible to produce a subset of
models that do give GRBs. This could be the small number of O and B
stars out on the tail of the rotational velocity distribution with
$v_{rot} \approx$ 400 km s$^{-1}$, which might be a few percent of all
such stars. It could equally well be a population of helium cores in
binaries that have arrived at helium ignition with a rotation rate
corresponding to about one-third of break up - provided the statistics
yield a sufficient number of such objects.  Either way, because of the
likely dependence of mass loss on metallicity \citep{Vin05}, GRBs will
be favored in regions of low metallicity as predicted by \citet{Mac99}
and observed by \citet{Pro04}.

\acknowledgments

At UCSC, this research has been supported by the NSF (AST 02-06111),
NASA (NAG5-12036), and the DOE Program for Scientific Discovery
through Advanced Computing (SciDAC; DE-FC02-01ER41176).  AH is
supported at LANL by DOE contract W-7405-ENG-36 to the Los Alamos
National Laboratory, and by NASA grants SWIF03-0047-0037 and NAG5-13700.

\clearpage

\onecolumn

\begin{deluxetable}{cccccccccc}
\setlength{\tabcolsep}{0.02in}
\tablecaption{Presupernova Models For Bare Helium Cores}
\label{tab:models}
\tablehead{ 
\colhead{Mass/} & 
\colhead{J$_{\rm initial}$} &
\colhead{Percent} & 
\colhead{$\dot M$} & 
\colhead{B-field} &
\colhead{M$_{\rm final}$} & 
\colhead{Fe-core} & 
\colhead{J$_{\rm Fe-core}$} &
\colhead{Period}& 
\colhead{a$_{\rm BH}$} \\ 
\colhead{Model} & 
\colhead{(10$^{52}$ erg s)} &
\colhead{break-up\tablenotemark{a}} &
\colhead{ WR } & 
\colhead{ } & 
\colhead{(M$_{\sun}$)} &
\colhead{(M$_{\sun}$)} & 
\colhead{(10$^{47}$ erg s)} & 
\colhead{(ms)} &
\colhead{(3 M$_{\sun}$)}} 
\startdata 
HE16A & 2.0 & 50 &  0.  &  no & 15.70 & 1.91 & 1870 & 0.06 & (6.0) \\
HE16B & 2.0 &  7 &  1.0 &  no &  5.10 & 1.45 &  275 & 0.27 & (1.0) \\
HE16C & 2.0 &  4 &  1.0 & yes &  5.15 & 1.44 &  5.8 & 12.0 & 0.02  \\
HE16D & 2.0 & 18 &  0.3 & yes &  9.53 & 1.61 & 33.4 &  2.4 & 0.16  \\
HE16E & 2.0 & 33 &  0.1 & yes & 12.86 & 1.38 & 45.4 &  1.4 & 0.23  \\
HE16F & 2.0 & 42 & 0.03 & yes & 14.80 & 1.75 &  114 & 0.86 & 0.48  \\
HE16G & 2.0 & 47 & 0.01 & yes & 15.56 & 1.90 &  146 & 0.82 & 0.51  \\
HE16H & 2.0 & 50 &  0.  & yes & 15.68 & 1.92 &  149 & 0.82 & 0.51  \\ 
      &     &    &      &     &       &      &      &      &       \\
HE16I & 1.0 & 26 &  0.  &  no & 15.88 & 1.89 & 1060 & 0.11 & (3.6) \\
HE16J & 1.0 &  5 &  1.0 &  no &  5.13 & 1.42 &  245 & 0.29 & 0.91  \\
HE16K & 1.0 &  2 &  1.0 & yes &  5.16 & 1.44 &  5.7 & 12.0 & 0.02  \\
HE16L & 1.0 & 10 &  0.3 & yes &  9.58 & 1.62 & 23.6 &  3.5 & 0.11  \\
HE16M & 1.0 & 17 &  0.1 & yes & 13.04 & 1.45 & 44.4 &  1.6 & 0.19  \\
HE16N & 1.0 & 22 & 0.03 & yes & 14.95 & 1.54 & 69.0 &  1.0 & 0.34  \\
HE16O & 1.0 & 24 & 0.01 & yes & 15.62 & 1.64 & 87.4 & 0.97 & 0.42  \\
HE16P & 1.0 & 26 & 0.   & yes & 15.88 & 1.83 &  119 & 0.91 & 0.45
\enddata 
\tablenotetext{a}{All models had a surface rotation rate of 47\% (Models A-H)
or 26\% (Models I-P) critical at helium ignition. The value given here is the
percentage of critical\ rotation at the surface when the star has
burned half its helium at its center.}
\end{deluxetable}

\begin{deluxetable}{ccccccccccc}
\setlength{\tabcolsep}{0.02in}
\tablecaption{Presupernova Models for Rapidly Rotating Stars}
\label{tab:models}
\tablehead{ 
\colhead{Mass/} & 
\colhead{J$_{\rm init}$\tablenotemark{a}} &
\colhead{v$_{\rm rot}$} & 
\colhead{PreSN}  &
\colhead{$\dot M$} &
\colhead{B-field}  &
\colhead{M$_{\rm final}$} &
\colhead{Fe-core} &
\colhead{J$_{\rm Fe-core}$} & 
\colhead{Period}&
\colhead{a$_{\rm BH}$} \\
\colhead{Model} & 
\colhead{(10$^{52}$ erg s)} &
\colhead{(km s$^{-1}$)} & 
\colhead{    } &
\colhead{ WR } &
\colhead{   }  &
\colhead{(M$_{\sun}$)} &
\colhead{(M$_{\sun}$)} &
\colhead{(10$^{47}$ erg s)} & 
\colhead{(ms)}   &
\colhead{(3 M$_{\sun}$)}}
\startdata 
12TA &   0.   &   0.  & RSG &       &  no  &  11.96  &  1.35  &   0.  &  -   &   -   \\
12TB &   1.5  & 290   & RSG &       &  no  &  11.95  &  1.42  &  362  & 0.17 & (2.4) \\
12TC &   2.0  & 380   & WR  &  0.1  &  no  &  11.56  &  1.57  & 1190  & 0.07 & (3.7) \\
12TD &   0.   &   0.  & RSG &       &  no  &  11.96  &  1.35  &   0.  &  -   &   -   \\
12TE &   1.5  & 290   & RSG &       &  no  &  11.95  &  1.43  &  369  & 0.17 & (2.4) \\ 
12TF &   2.0  & 380   & WR  &  1.0  &  no  &   9.18  &  1.50  &  606  & 0.13 & (2.1) \\
12TG &   1.5  & 290   & RSG &       & yes  &  11.96  &  1.41  &  6.2  & 9.5  & 0.06  \\
12TH &   2.0  & 380   & WR  &  1.0  & yes  &   9.23  &  1.48  &  43.2 & 1.8  & 0.17  \\
12TI &   2.0  & 380   & WR  &  0.3  & yes  &  10.79  &  1.57  &  65.0 & 1.2  & 0.31  \\
12TJ &   2.0  & 380   & WR  &  0.1  & yes  &  11.54  &  1.82  & 150   & 0.72 & 0.57  \\
     &        &       &     &       &      &         &        &       &      &       \\
16TA &   0.   &   0.  & BSG &       &  no  &  15.95  &  1.44  &   0.  &  -   &   -   \\
16TB &   2.5  & 305   & WR  &  0.1  &  no  &  15.29  &  1.75  & 1640  & 0.06 & (5.7) \\
16TC &   3.3  & 390   & WR  &  0.1  &  no  &  15.23  &  1.61  & 1420  & 0.06 & (6.0) \\
16TD &   0.   &   0.  & BSG &       &  no  &  15.95  &  1.44  &   0.  &  -   &   -   \\
16TE &   2.5  & 305   & WR  &  1.0  &  no  &  11.98  &  1.47  &  459  & 0.17 & (1.4) \\
16TF &   3.3  & 390   & WR  &  1.0  &  no  &  11.37  &  1.85  &  596  & 0.18 & (1.6) \\
16TG &   2.5  & 305   & RSG &       & yes  &  15.67  &  1.75  &  19.2 & 5.1  & 0.05  \\
16TH &   3.3  & 390   & WR  &  1.0  & yes  &  11.45  &  1.81  &  64.1 & 1.7  & 0.24  \\
16TI &   3.3  & 390   & WR  &  0.3  & yes  &  13.95  &  1.60  &  86.7 & 0.90 & 0.44  \\
16TJ &   3.3  & 390   & WR  &  0.1  & yes  &  15.21  &  1.88  & 178   & 0.67 & 0.61  \\
     &        &       &     &       &      &         &        &       &      &       \\
12OA &    0.  &   0.  & RSG &       &  no  &  11.86  &  1.25  &   0.  &  -   &  -    \\
12OB &   1.5  & 245   & RSG &       &  no  &  11.83  &  1.43  &  345  & 0.18 & (2.2) \\
12OC &   2.0  & 325   & WR  &  0.1  &  no  &  10.98  &  1.57  &  1120 & 0.07 & (5.2) \\
12OD &    0.  &   0.  & RSG &       &  no  &  11.86  &  1.32  &   0.  &  -   &  -    \\
12OE &   1.5  & 245   & RSG &       &  no  &  11.83  &  1.43  &  343  & 0.18 & (2.2) \\
12OF &   2.0  & 325   & WR  &  1.0  &  no  &   7.61  &  1.42  &  354  & 0.18 & (1.2) \\
12OG &   1.5  & 245   & RSG &       & yes  &  11.83  &  1.41  &  6.1  & 9.7  & 0.06  \\
12OH &   2.0  & 325   & WR  &  1.0  & yes  &   7.69  &  1.39  &  10.0 & 6.1  & 0.04  \\
12OI &   2.0  & 325   & WR  &  0.3  & yes  &   9.71  &  1.41  &  39.5 & 1.6  & 0.17  \\
12OJ &   2.5  & 400   & WR  &  0.1  &  no  &  10.95  &  1.52  & 1160  & 0.07 & (5.0) \\  
12OK &   2.5  & 400   & WR  &  1.0  &  no  &   7.30  &  1.35  &  314  & 0.18 & (1.2) \\
12OL &   2.5  & 400   & WR  &  1.0  & yes  &   7.35  &  1.35  &  9.5  & 5.8  & 0.04  \\
12OM &   2.5  & 400   & WR  &  0.3  & yes  &   9.50  &  1.86  & 109   & 1.0  & 0.39  \\
12ON &   2.5  & 400   & WR  &  0.1  & yes  &  10.93  &  1.50  &  74.1 & 1.0  & 0.44  \\
     &        &       &     &       &      &         &        &       &      &       \\
16OA &    0.  &   0.  & RSG &       &  no  &  15.80  &  1.43  &   0.  &  -   &  -    \\
16OB &   2.5  & 255   & RSG &       &  no  &  15.57  &  1.84  &  523  & 0.20 & (1.8) \\
16OC &   3.3  & 325   & WR  &  0.1  &  no  &  14.26  &  1.63  &  1360 & 0.06 & (4.4) \\
16OD &    0.  &   0.  & RSG &       &  no  &  15.80  &  1.43  &   0.  &  -   &  -    \\
16OE &   2.5  & 255   & RSG &       &  no  &  15.57  &  1.84  &  523  & 0.20 & (1.8) \\
16OF &   3.3  & 325   & WR  &  1.0  &  no  &   8.97  &  1.35  &  318  & 0.18 & (1.1) \\
16OG &   2.5  & 255   & RSG &       & yes  &  15.66  &  1.50  &  9.6  & 7.0  & 0.05  \\
16OH &   3.3  & 325   & WR  &  1.0  & yes  &   9.18  &  1.45  &  9.8  & 7.9  & 0.03  \\
16OI &   3.3  & 325   & WR  &  0.3  & yes  &  12.21  &  1.65  &  55.3 & 1.5  & 0.26  \\
16OJ &   4.1  & 400   & WR  &  0.1  &  no  &  14.20  &  1.56  & 1290  & 0.06 & (5.0) \\
16OK &   4.1  & 400   & WR  &  1.0  &  no  &   8.58  &  1.52  &  399  & 0.21 & (1.4) \\
16OL &   4.1  & 400   & WR  &  1.0  & yes  &   8.68  &  1.52  &  14.9 & 5.6  & 0.05  \\
16OM &   4.1  & 400   & WR  &  0.3  & yes  &  11.94  &  1.55  &  53.3 & 1.4  & 0.25  \\
16ON &   4.1  & 400   & WR  &  0.1  & yes  &  14.17  &  1.78  &  121  & 0.85 & 0.43  \\
     &        &       &     &       &      &         &        &       &      &       \\
12SA &    0.  &  0.   & RSG &       &  no  &  10.88  &  1.40  &  0.   &  -   &  -    \\
12SB &   2.0  & 280   & WR  &  0.1  &  no  &   9.39  &  1.40  &  419  & 0.16 & (1.7) \\
12SC &   3.0  & 405   & WR  &  0.1  &  no  &   8.92  &  1.66  &  751  & 0.11 & (2.8) \\
12SD &    0.  &  0.   & RSG &       &  no  &  10.88  &  1.40  &  0.   &  -   &  -    \\
12SE &   2.0  & 280   & WR  &  1.0  &  no  &   7.27  &  1.42  &  355  & 0.19 & (1.1) \\
12SF &   3.0  & 405   & WR  &  1.0  &  no  &   4.96  &  1.51  &  333  & 0.23 & (1.1) \\
12SG &   2.0  & 280   & BSG &       & yes  &   7.57  &  1.57  &  9.4  & 8.3  & 0.04  \\
12SH &   3.0  & 405   & WR  &  1.0  & yes  &   5.43  &  1.46  &  6.6  & 11   & 0.03  \\
12SI &   3.0  & 405   & WR  &  0.3  & yes  &   6.95  &  1.39  &  7.1  & 8.9  & 0.03  \\
12SJ &   3.0  & 405   & WR  &  0.1  & yes  &   9.03  &  1.46  & 36.6  & 1.7  & 0.15  \\
     &        &       &     &       &      &         &        &       &      &       \\
16SA &    0.  &  0.   & RSG &       &  no  &  14.65  &  1.38  &  0.   &  -   &  -    \\
16SB &   2.5  & 215   & RSG &       &  no  &  14.03  &  1.44  &  332  & 0.23 & (1.2) \\
16SC &   3.3  & 270   & WR  &  0.1  &  no  &  11.58  &  1.73  &  492  & 0.19 & (1.4) \\
16SD &    0.  &       & RSG &       &  no  &  14.65  &  1.38  &  0.   &  -   &  -    \\
16SE &   2.5  & 215   & RSG &       &  no  &  14.03  &  1.44  &  332  & 0.23 & (1.2) \\
16SF &   3.3  & 270   & WR  &  1.0  &  no  &   6.25  &  1.35  &  281  & 0.20 & (1.1) \\
16SG &   2.5  & 215   & RSG &       & yes  &  11.98  &  1.49  &  9.5  & 7.9  & 0.04  \\
16SH &   3.3  & 270   & WR  &  1.0  & yes  &   7.70  &  1.35  &  8.2  & 6.7  & 0.03  \\
16SI &   3.3  & 270   & WR  &  0.3  & yes  &   9.85  &  1.72  &  14.8 & 6.4  & 0.06  \\  
16SI &   4.5  & 360   & WR  &  0.1  &  no  &  11.10  &  1.76  &  550  & 0.18 & (1.6) \\
16SJ &   4.5  & 360   & WR  &  1.0  &  no  &   5.33  &  1.46  &  293  & 0.25 & (1.0) \\
16SK &   4.5  & 360   & WR  &  1.0  & yes  &   6.30  &  1.33  &  5.9  & 9.7  & 0.02  \\
16SL &   4.5  & 360   & WR  &  0.3  & yes  &   8.31  &  1.52  &  10.9 & 7.6  & 0.04  \\
16SM &   4.5  & 360   & WR  &  0.1  & yes  &  11.22  &  1.90  &  50.4 & 2.3  & 0.16  \\
     &        &       &     &       &      &         &        &       &      &       \\
35OA &   14.  & 380   & WR  &  1.0  & yes  &  12.86  &  1.52  &  13.2 & 6.0  & 0.05  \\
35OB &   14.  & 380   & WR  &  0.3  & yes  &  21.24  &  2.05  &  151  & 0.86 & 0.38  \\
35OC &   14.  & 380   & WR  &  0.1  & yes  &  28.07  &  2.02  &  230  & 0.59 & 0.53  \\
35OA &   14.  & 380   & WR  &  0    & yes  &  34.4   &  1.29  &  193  & 0.28 & (1.2) \\
\enddata
\tablenotetext{a}{Total angular momentum of the zero age main sequence star.}  
\end{deluxetable}

\clearpage 
\begin{figure}
\newcommand{\panelwidth}{0.48\columnwidth}
\includegraphics[draft=\Draft,angle=0,width=\panelwidth]{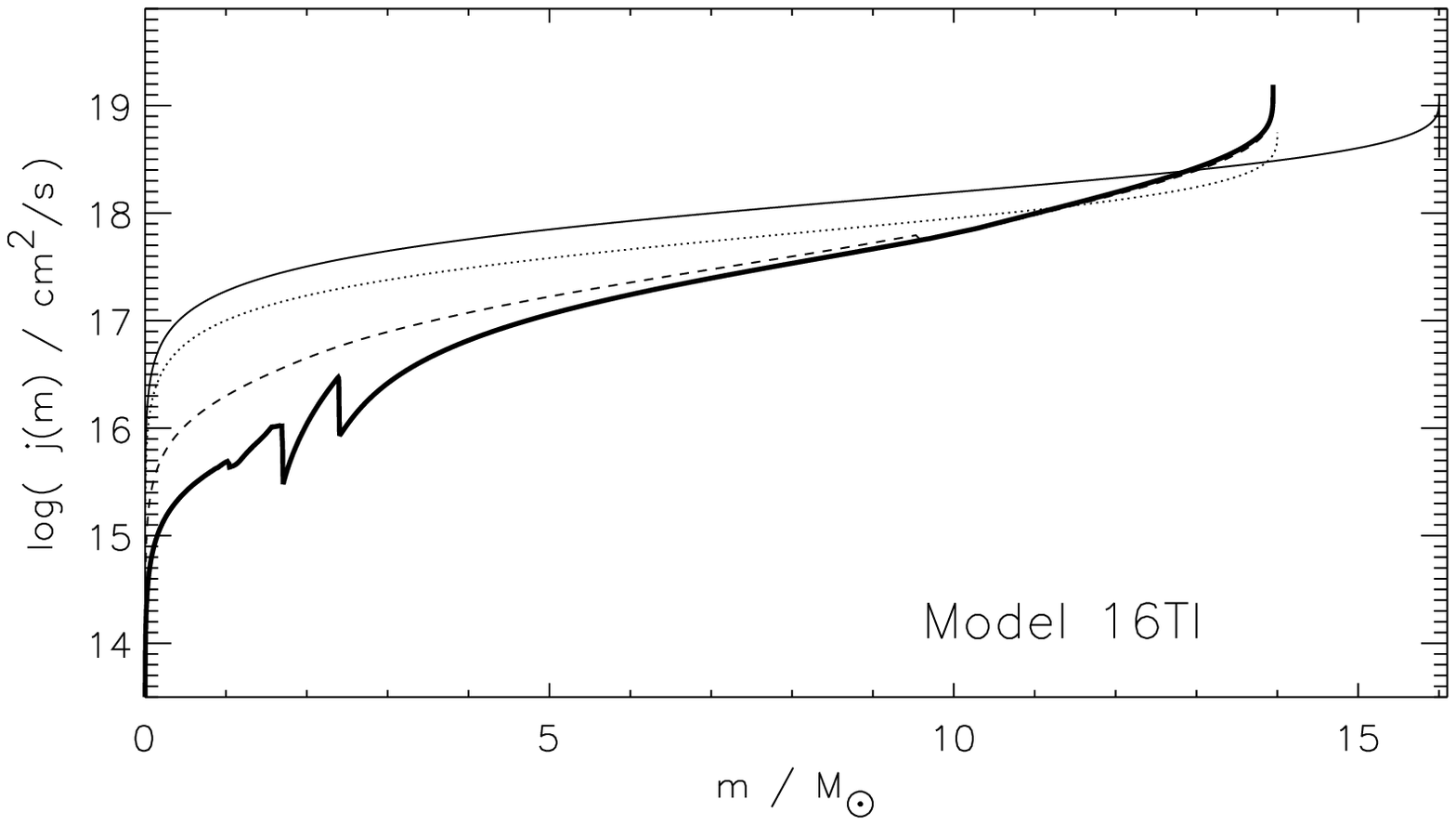}\hfill
\includegraphics[draft=\Draft,angle=0,width=\panelwidth]{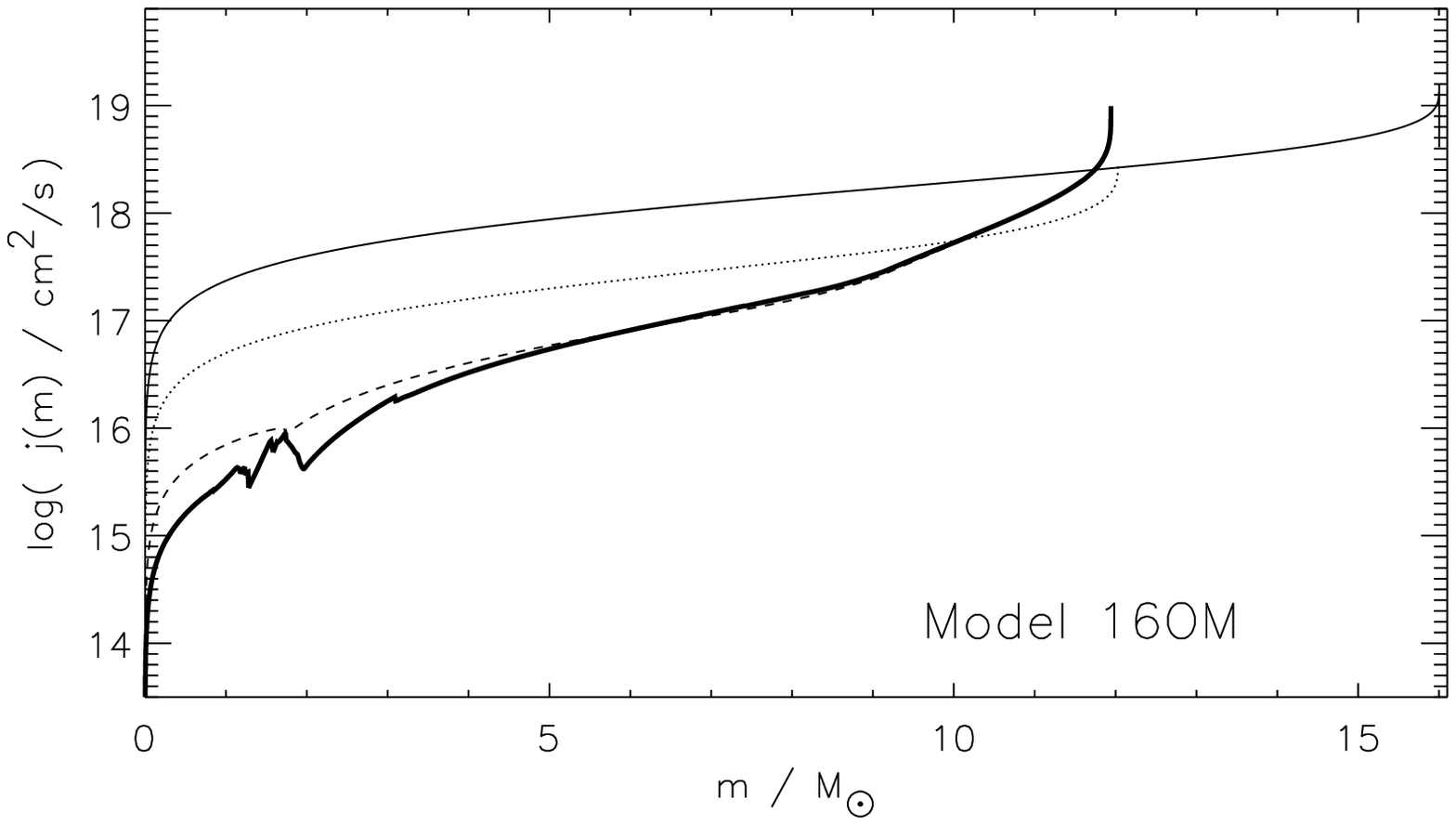}\\[\baselineskip]
\includegraphics[draft=\Draft,angle=0,width=\panelwidth]{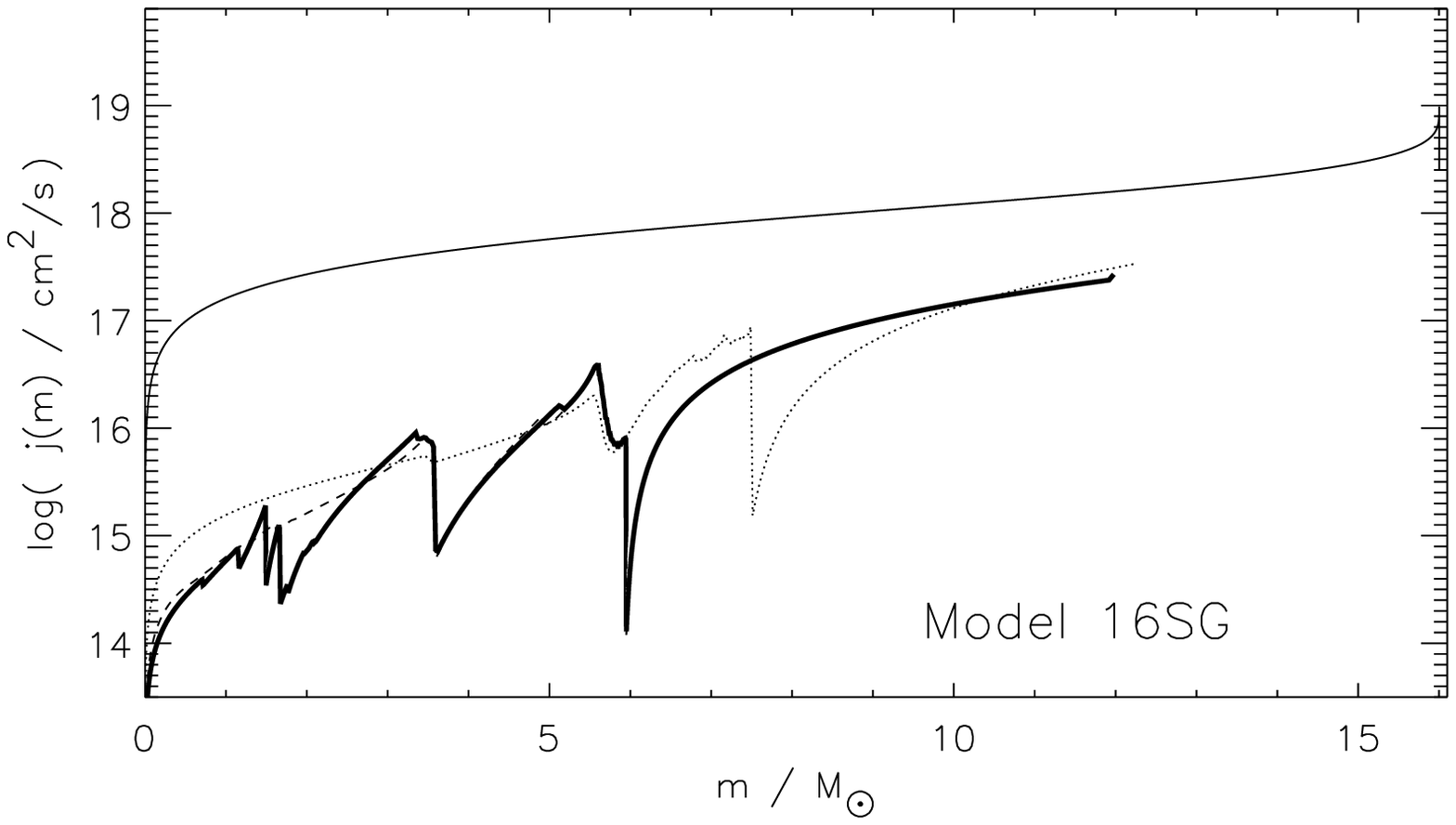}\hfill
\includegraphics[draft=\Draft,angle=0,width=\panelwidth]{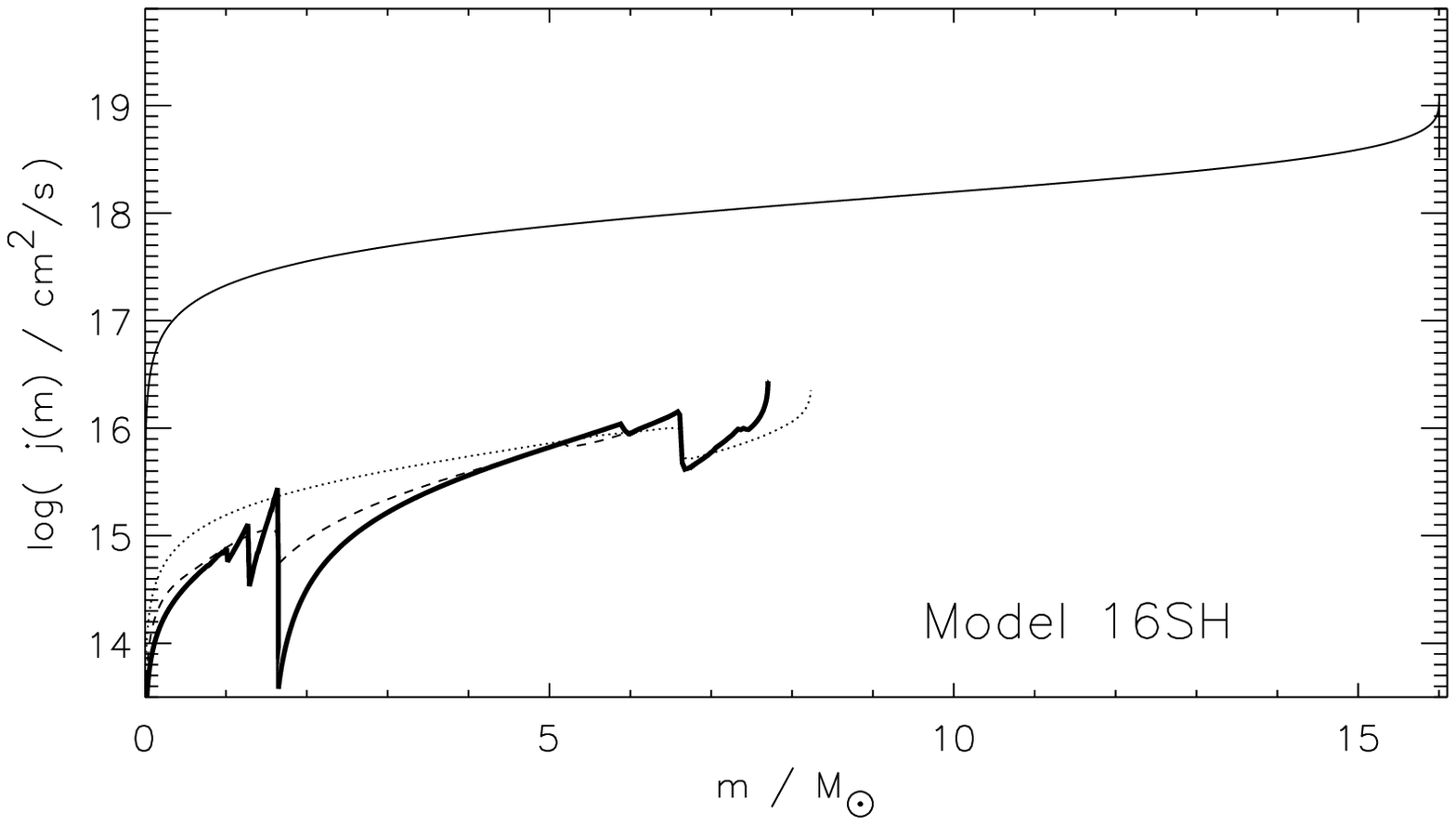}
\caption{Distribution of the specific angular momentum with mass in
four 16 solar mass models evaluated at the zero age main sequence (top
line, solid); central helium depletion (dotted line, next down);
carbon depletion (dashed line, second down); and at the presupernova
stage (dark solid line, bottommost). The upper two models had reduced
metallicity (1\% solar for the left frame; 10\% for the right) and
mass loss during the WR stage (30\%). Both rotated at about 400 km
s$^{-1}$ on the main sequence and avoided forming a red giant. The
models in the bottom two panels had solar metallicity and more
moderate rotation on the main sequence. The one on the right rotated
faster and ended up a WR star. The one on the left died as a red
supergiant. The pulsars derived from the top two models rotated 5 to
10 times faster. Avoiding the formation of a giant star and suppressing
mass loss amplifies the final rotational momentum of the core.
\lFig{jevol}}
\end{figure}

\clearpage 
\begin{figure} 
\centering 
\includegraphics[draft=\Draft,angle=0,width=\columnwidth]{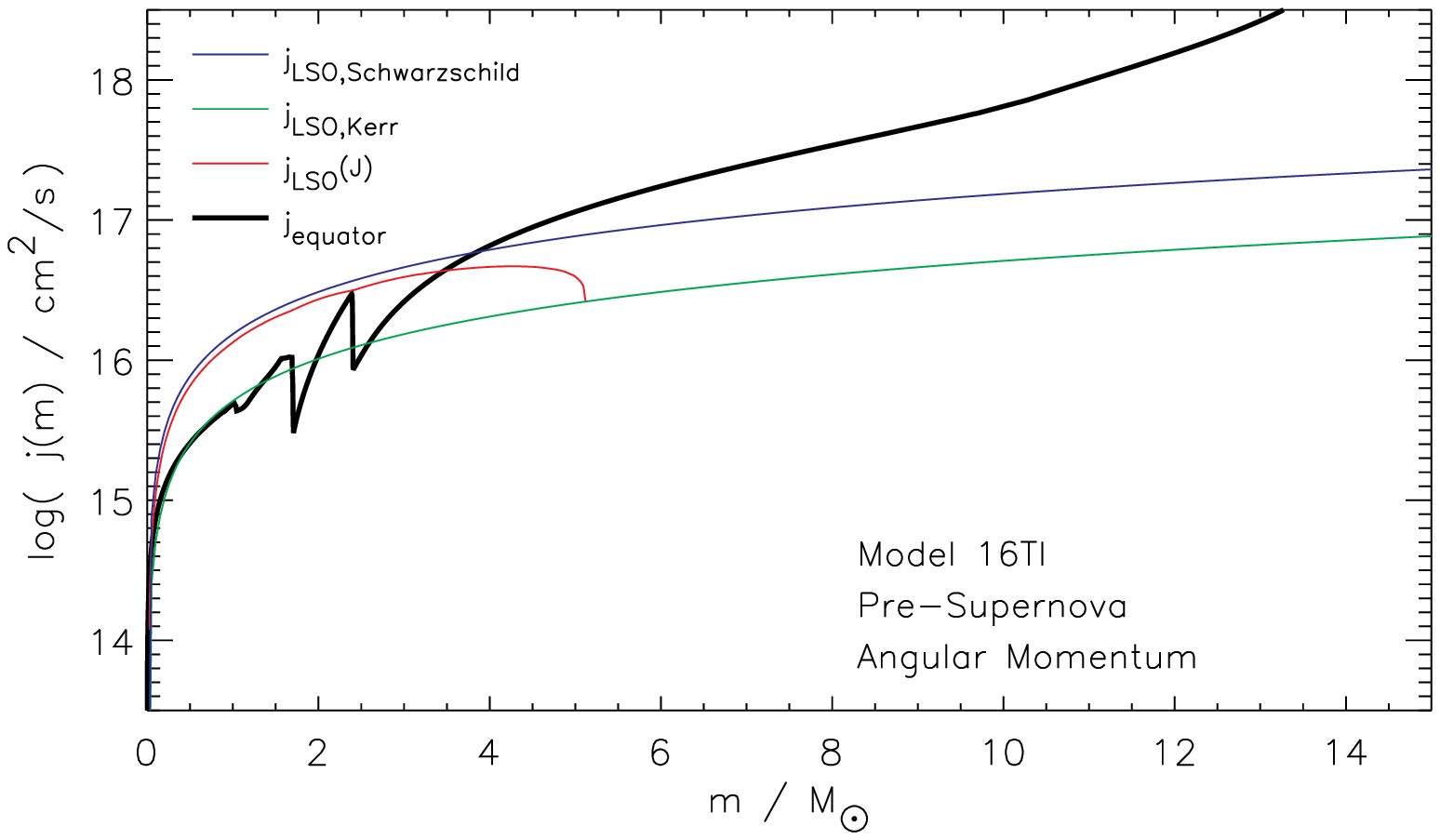}
\includegraphics[draft=\Draft,angle=0,width=\columnwidth]{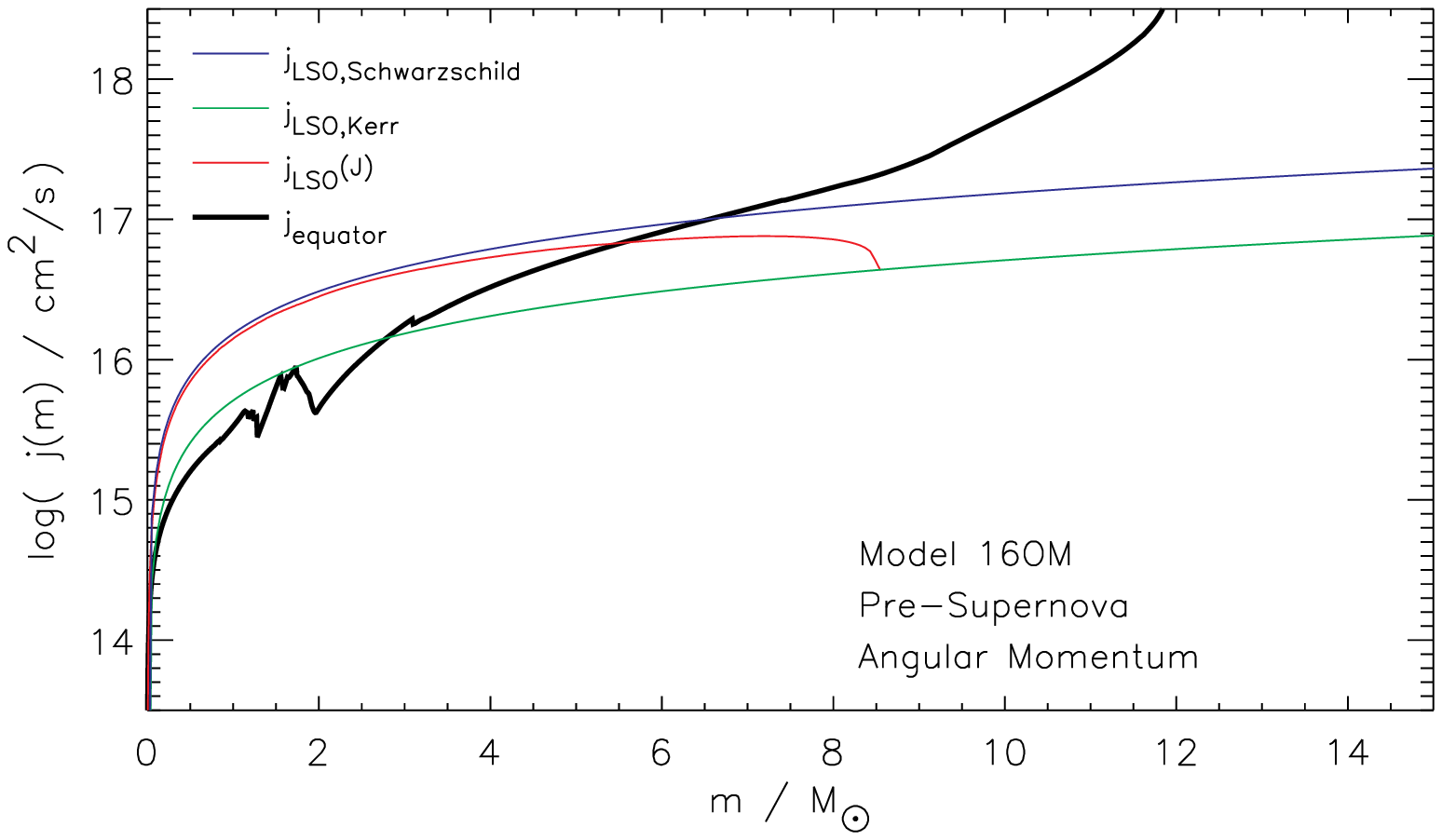}
\caption{Angular momentum distribution of Models 16TI and 16OM at the
time the star collapsed. The dark solid line shows the distribution of
specific angular momentum in the presupernova star. The blue line
indicates the angular momentum required to support matter at the
stable orbit for a black hole that is not rotating; the green line is
for a Kerr black hole with rotational parameter a = 1. The red line
indicates the last stable orbit for a black hole with the mass and
angular momentum inside the indicated coordinate in the presupernova
star. Where the black line is above the red line a disk can form and
collapsars are possible.
\lFig{jpresn}}
\end{figure}


\begin{thebibliography}{99}

\bibitem[Abt et al.(2002)]{Abt02} 
Abt, H.~A., Levato, H., \& Grosso, M.\ 2002, \apj, 573, 359

\bibitem[Arras et al.(2003)]{Arr03} 
Arras, P., Flanagan, E.~E., Morsink, S.~M., Schenk, A.~K., Teukolsky,
S.~A., \& Wasserman, I.\ 2003, \apj, 591, 1129
 
\bibitem[Berger et al.(2002)]{Ber02} 
Berger, E., Kulkarni, S.~R., \& Chevalier, R.~A.\ 2002, \apjl, 577, L5

\bibitem[Brown et al.(2004)]{Bro04} 
Brown, J.~C., Cassinelli, J.~P., Li, Q., Kholtygin, A.~F., \& Ignace,
R.\ 2004, \aap, 426, 323

\bibitem[Crowther et al.(2002)]{Cro02} 
Crowther, P.~A., Dessart, L., Hillier, D.~J., Abbott, J.~B., \&
Fullerton, A.~W.\ 2002, \aap, 392, 653

\bibitem[Deng et al.(2005)]{Den05} 
Deng, J., Tominaga, N., Mazzali, P.~A., Maeda, K., \& Nomoto, K.\
2005, \apj, 624, 898
 
\bibitem[Fruchter et al.(1999)]{Fru99} 
Fruchter, A.~S., et al.\ 1999, \apjl, 519, L13

\bibitem[Fryer \& Heger(2005)]{Fry05} 
Fryer, C.~L., \& Heger, A.\ 2005, \apj, 623, 302
 
\bibitem[Galama et al.(1998)]{Gal98} 
Galama, T.~J., et al.\ 1998, \nat, 395, 670
  
\bibitem[Gies \& Huang(2005)]{Gie05}
Gies, D. R., \& Huang, W. 2005, in {\sl Stellar Rotation}, IAU
Symposium 215, ed. A. Maeder and P. Eenens, ASP, p. 57

\bibitem[Hamann \& Koesterke(1998)]{HK98}
Hamann, W.-R., Koesterke, L.\ 1998, \aap, 335, 1003

\bibitem[Hanuschik(1996)]{Han96} 
Hanuschik, R.~W.\ 1996, \aap, 308, 170

\bibitem[Heger et al.(2000)]{Heg00} 
Heger, A., Langer, N., \& Woosley, S.~E.\ 2000, \apj, 528, 368

\bibitem[Heger \& Woosley(2003)]{Heg03} 
Heger, A., \& Woosley, S.~E.\ 2003, AIP Conf.~Proc.~662: Gamma-Ray
Burst and Afterglow Astronomy 2001: A Workshop Celebrating the First
Year of the HETE Mission, 662, 214
 

\bibitem[Heger et al.(2005)]{Heg05} 
Heger, A., Woosley, S.~E., \& Spruit, H.~C.\ 2005, \apj, 626, 350

\bibitem[Hirschi et al.(2004)]{Hir04} 
Hirschi, R., Meynet, G., \& Maeder, A.\ 2004, \aap, 425, 649

\bibitem[Hirschi et al.(2005)]{Hir05} 
Hirschi, R., Meynet, G., \& Maeder, A.\ 2005, \aap, in press
 
\bibitem[Hjorth et al.(2002)]{Hjo02} 
Hjorth, J., et al.\ 2002, \apj, 576, 113

\bibitem[Iwamoto et al.(1998)]{Iwa98} 
Iwamoto, K., et al.\ 1998, \nat, 395, 672
 
\bibitem[Kudritzki(2000)]{Kud00}
Kudritzki, R.-P.\ 2000,
Proceedings of the MPA/ESO Workshop on ``The First Stars'', 
eds.\ A.\ Weiss, T.~G.\ Abel, V.\ Hill, Springer, p.~127

\bibitem[Kudritzki(2002)]{Kud02}
Kudritzki, R.~P.\ 2002, \apj, 577, 389

\bibitem[Langer(1989)]{Lan89}
Langer, N. 1989, \aap, 220, 135

\bibitem[Lattimer \& Prakash(2001)]{Lat01}
Lattimer, J.~M., \& Prakash, M.\ 2001, \apj, 550, 426
 
\bibitem[Levan et al.(2005)]{Lev05} 
Levan, A., et al.\ 2005, \apj, 624, 880

\bibitem[Lodders(2003)]{Lod03} 
Lodders, K.\ 2003, \apj, 591, 1220
 
\bibitem[MacFadyen \& Woosley(1999)]{Mac99} 
MacFadyen, A.~I., \& Woosley, S.~E.\ 1999, \apj, 524, 262
 
\bibitem[Maeder(2002)]{Mae02} 
Maeder, A.\ 2002, \aap, 392, 575

\bibitem[Meynet \& Maeder(2005)]{Mey05} 
Meynet, G., \& Maeder, A.\ 2005, \aap, 429, 581

\bibitem[Nieuwenhuijzen \& de Jager(1990)]{NJ90}
Nieuwenhuijzen, H., de Jager, C.\ 1990, \aap, 231, 134
 
\bibitem[Nugis et al.(1998)]{Nu98} 
Nugis, T., Crowther, P.~A., \& Willis, A.~J.\ 1998, \aap, 333, 956
 
\bibitem[Nugis \& Lamers(2000)]{NL00}
Nugis, T., \& Lamers, H.J.G.L.M.\ 2000, \aap, 360, 227

\bibitem[Soderberg et al.(2005)]{Sod05} Soderberg, A.~M., 
Kulkarni, S.~R., Berger, E., Chevalier, R.~A., Frail, D.~A., Fox, D.~B., \& 
Walker, R.~C.\ 2005, \apj, 621, 908 

\bibitem[Petrovic et al.(2005)]{Pet05} 
Petrovic, J., Langer, N., Yoon, S.-C., \& Heger, A.\ 2005, \aap, 435,
247

\bibitem[Podsiadlowski et al.(2004)]{Pod04} 
Podsiadlowski, P., Mazzali, P.~A., Nomoto, K., Lazzati, D., \&
Cappellaro, E.\ 2004, \apjl, 607, L17

\bibitem[Prochaska et al.(2004)]{Pro04} 
Prochaska, J.~X., et al.\ 2004, \apj, 611, 200

\bibitem[Smartt et al.(2002)]{Sma02} 
Smartt, S.~J., Vreeswijk, P.~M., Ramirez-Ruiz, E., Gilmore, G.~F.,
Meikle, W.~P.~S., Ferguson, A.~M.~N., \& Knapen, J.~H.\ 2002, \apjl,
572, L147
  
\bibitem[Spruit(2002)]{Spr02} 
Spruit, H.~C.\ 2002, \aap, 381, 923
 
\bibitem[Stanek et al.(2003)]{Sta03} 
Stanek, K.~Z., et al.\ 2003, \apjl, 591, L17

\bibitem[Timmes et al.(1996)]{Tim96} 
Timmes, F.~X., Woosley, S.~E., \& Weaver, T.~A.\ 1996, \apj, 457, 834
 
\bibitem[Townsend et al.(2004)]{Tow04} 
Townsend, R.~H.~D., Owocki, S.~P., \& Howarth, I.~D.\ 2004, \mnras,
350, 189

\bibitem[Usov(1994)]{Uso94} 
Usov, V.~V.\ 1994, \mnras, 267, 1035
 
\bibitem[Vanbeveren(2001)]{Van01}
Vanbeveren, D.\ 2001, in \textsl{The Influence of Binaries on Stellar 
Population Studies}, ASSL series,
ed.\ D.\ Vanbeveren, Kluwer Publishers; astroph-0201110


\bibitem[Vietri \& Stella(1998)]{Vie98} 
Vietri, M., \& Stella, L.\ 1998, \apjl, 507, L45

\bibitem[Vink \& de Koter(2005)]{Vin05}
Vink, J. S., \& de Koter, A. 2005, submitted to \aap, astroph 0507352
 
\bibitem[Vreeswijk et al.(2001)]{Vre01} 
Vreeswijk, P.~M., et al.\ 2001, \apj, 546, 672
 
\bibitem[Weaver et al.(1978)Weaver, Zimmerman, \& Woosley]{Wea78}
Weaver, T. A., Zimmerman, G. B., \& Woosley, S. E. 1978, \apj, 225, 1021

\bibitem[Wellstein \& Langer(1999)]{WL99}
Wellstein, S., Langer, N.\ 1999, \aap, 350, 148

\bibitem[Wheeler et al.(2000)]{Whe00}
Wheeler, J. C., Yi, I., H\"oflich, P., \& Wang, L. 2000, \apj, 
537, 810

\bibitem[Woosley(1993)]{Woo93} 
Woosley, S.~E.\ 1993, \apj, 405, 273

\bibitem[Woosley et al.(1999)]{Woo99} 
Woosley, S.~E., Eastman, R.~G., \& Schmidt, B.~P.\ 1999, \apj, 516,
788

\bibitem[Woosley et al.(2002)Woosley, Heger, Weaver]{WHW02} 
Woosley, S.~E., Heger, A., Weaver, T.~A.\ 2002, Rev.\ Mod.\ Phys., 
in press

\bibitem[Woosley \& Heger(2004)]{Woo04} 
Woosley, S.~E., \& Heger, A.\ 2004, IAU Symposium, 215, 601

\bibitem[Zeh et al.(2004)]{Zeh04} 
Zeh, A., Klose, S., \& Hartmann, D.~H.\ 2004, \apj, 609, 952

\bibitem[Zhang et al.(2004)]{Zha04} 
Zhang, W., Woosley, S.~E., \& Heger, A.\ 2004, \apj, 608, 365
 
 
\end{thebibliography}
\end{document}